\documentclass[conference]{IEEEtran}
\usepackage{cite}
\usepackage{amsmath,amssymb,amsfonts}
\usepackage{algorithmic}
\usepackage{graphicx}
\usepackage{textcomp}
\usepackage{xcolor}

\def\BibTeX{{\rm B\kern-.05em{\sc i\kern-.025em b}\kern-.08em
    T\kern-.1667em\lower.7ex\hbox{E}\kern-.125emX}}

\newcommand\blankfootnote[1]{%
	\let\thefootnote\relax\footnotetext{#1}%
	\let\thefootnote\svthefootnote%
}

\begin{document}

\title{An Exploration of the Mentorship Needs of Research Software Engineers}

\author{\IEEEauthorblockN{Reed Milewicz}
\IEEEauthorblockA{\textit{Department of Software Engineering and Research} \\
\textit{Sandia National Laboratories}\\
Albuquerque, NM \\
rmilewi@sandia.gov}
\and
\IEEEauthorblockN{Miranda Mundt}
\IEEEauthorblockA{\textit{Department of Software Engineering and Research} \\
\textit{Sandia National Laboratories}\\
Albuquerque, NM \\
mmundt@sandia.gov}}

\maketitle

\begin{abstract}
As a newly designated professional title, research software engineers (RSEs) link the two worlds of software engineering and research science. They lack clear development and training opportunities, particularly in the realm of mentoring. In this paper, we discuss mentorship as it pertains to the unique needs of RSEs and propose ways in which organizations and institutions can support mentor/mentee relationships for RSEs.
\end{abstract}

\blankfootnote{Sandia National Laboratories is a multimission laboratory managed and operated by National Technology \& Engineering Solutions of Sandia, LLC, a wholly owned subsidiary of Honeywell International Inc., for the U.S. Department of Energy’s National Nuclear Security Administration under contract DENA0003525. SAND2021-12402 C. Presented at Research Software Engineers in HPC (RSE-HPC-2021), colocated with Supercomputing'21 (SC'21). }


\section{Introduction}

The past decade has seen the rise of the research software engineer (RSE) as a distinct professional identity for those advancing software engineering practice within the scientific domain~\cite{baxter2012}. The RSE movement has flourished~\cite{brett2017research,cohen2020four}, leading to the creation of official RSE departments at national labs  and universities~\cite{katz2019research,milewicz2020research,billings2020}, the emergence of national, non-profit RSE organizations~\cite{deRSEWebsite,nlRSEWebsite,usSEWebsite,nzRSEWebsite,ukRSEWebsite}, and the establishment of multiple conferences, workshops, and other colloquia~\cite{se4scienceWorkshops,usrseWorkshops,rseconWorkshops}. As we look to the future, sustaining that growth by promoting the skill development and career advancement of RSEs will be a priority. Currently, there are no formal education programs for RSEs as there are for conventional software engineers or researchers; as a rule, RSEs come from a diverse variety of backgrounds with different competencies and learning needs. As we move into the exascale era and beyond, we anticipate many more disruptive cycles of innovation in computing technologies  -- RSEs will need to remain current with knowledge across multiple disciplines to meet the demand for high quality software. Moreover, we must also consider how to retain talented RSEs by giving more recognition to their work, engendering a shared sense of professional identity, and cultivating a community of practice. 

In this paper, we examine mentorship as a strategy to enable the career growth and retention of RSEs. Mentorship is widely recognized as a vehicle for professional development in numerous fields and is likely to be beneficial for RSEs. However, it is not yet understood what this mentorship ought to look like or how teams and organizations should promote mentoring activities. To that end, we (1) provide a brief overview of the literature on mentorship, (2) identify key needs that RSEs may have in providing or receiving mentorship, and finally (3) propose directions for future work.

\section{Background}
\label{sec:background}

By mentorship, we mean a relationship in which a more experienced or more knowledgeable person (a mentor) helps to guide a less experienced or less knowledgeable person (a mentee). Mentorship is seen as an essential component of career development in fields such as teaching~\cite{hobson2009mentoring}, academic medicine~\cite{sambunjak2010systematic}, and nursing~\cite{andrews1999mentorship}, and it has likewise been studied in the contexts of software development~\cite{trainer2017mentoring,canfora2012going,fagerholm2014role} and science and mathematics~\cite{lienard2018intellectual,malmgren2010role}. A full treatment of the scholarship around mentorship is beyond the scope of this paper, but it is widely known that, for mentees, the mentoring process can be a valuable mechanism for both career support (i.e., training, coaching, and advocacy) and psychosocial support (i.e., role modeling, counseling, and friendship)~\cite{kram1983phases}; for mentors, mentorship can provide personal satisfaction, create opportunities for learning through teaching, and reinforce their sense of professional identity (e.g., \cite{wright1997perceptions}). In short, promoting mentorship could prove valuable for RSE teams and organizations.

\section{The Mentorship Needs of RSEs}

There is currently a lack of specific, evidence-based guidance for teams and institutions looking to support the mentorship of RSEs. To promote dialogue and spur future work in this direction, we have identified three key areas -- based on the available literature and on our own experiences -- where we believe RSEs have distinct needs that mentorship programs could help address.

\textbf{Interdisciplinary Mentorship Networks}: A unique trait of RSEs is their interdisciplinary backgrounds. An RSE may not have received formalized software engineering education or training; rather, she may have studied a different STEM field, such as mathematics or chemical engineering, and has subsequent understanding of that particular research domain~\cite{baxter2012}. Even given a traditional software engineering education, universities often lag behind the needs of industry~\cite{jazayeri2004} and rarely provide the specific knowledge and skills needed for scientific computing. If future RSEs are unable to get proper education both in their preferred STEM domain and software engineering, mentorship must play a larger role. Given the interdisciplinary character of RSE work, however, a unique requirement for effective RSE mentorship is the availability of multiple mentors across different domains. No single mentor will be able to supply the necessary organizational, engineering, and domain expertise to a new RSE. As De Janasz et al. have noted, mentees in general benefit from access to a network of mentors who can provide different kinds of learning opportunities (namely learning the "why", "how", and "who" of their field)~\cite{de2003}. This is especially true, however, for professionals in interdisciplinary domains~\cite{henry2019mentorship,guise2017team,freel2017multidisciplinary,kensington2018surviving}, such as RSEs; to borrow phrasing from Jazayeri, ``an interdisciplinary software engineer must be conversant in other disciplines''\cite{jazayeri2004}. Building an effective network of willing software engineers, domain experts, and organizational culture mentors will boost an RSE's confidence, skill set, and career growth opportunities.

\textbf{Long-term Mentor/Mentee Relationships}: Today it is common for developers in industry to frequently change jobs in pursuit of growth opportunities and better salaries. While a 2010 study found that one-third of new hires across all industries were expected to leave an organization within 2 years of hire~\cite{stein2010}, this trend is especially pronounced in software engineering where, as of 2016, the median tenure of developers at major tech firms ranged from 1.5 to 2.3 years~\cite{dice2016}. This is radically different from the workplace culture of most universities and national labs where professors and staff scientists may remain with the same institution for decades. If we intend to retain RSEs, we need a mentoring model that facilitates long-term, ongoing career growth. RSEs are constantly transforming -- they must stay apprised of changes in technology, industry, and science in order to remain relevant within their field. Whereas mentorship in conventional software engineering is frequently limited to onboarding, mentorship should not stop once an RSE shows a certain level of independence but instead needs to remain a high-level priority throughout the career of any RSE. 

\textbf{Training Soft Skills}: Software engineering researchers and practitioners have called attention to the importance of soft skills in software development such as teamwork, communication, and leadership skills~\cite{matturro2015soft,capretz2018call,matturro2019systematic}. Soft skills are especially relevant for RSEs who act as a bridge between worlds and as an interpreter between cultures. RSEs must communicate with domain experts, navigate research institutions as a software professional, and articulate software engineering best practices in the scientific domain. Unfortunately, RSEs rarely (if ever) receive any formal training on applying those kinds of interpersonal skills, and an explicit goal of RSE mentorship should be to fill that gap. As we mentioned in Section \ref{sec:background}, psychosocial support is a critical function of mentorship, and this can include counseling mentees on relationships with colleagues, encouraging self-reflection, and lining up work opportunities to exercise soft skills. 

\section{Discussion}

Turning our attention to immediate and practical considerations, providing RSEs with mentorship opportunities will require explicit institutional support. Many RSEs at universities and national laboratories are matrixed into multiple concurrent research software projects\cite{katz2019research}, and many RSEs play critical roles in small projects that demand their attention~\cite{van2018survey}. With such a packed and diverse project load, there will likely be a lack of time and funding to participate in mentorship activities. Institutions can help this effort by specifically carving out overhead funds to support mentor/mentee activities and relationships.  This can include formal mentorship programs, tailored training for RSEs who want to offer mentorship, incentives for engaging in mentorship, and protected time for mentors and mentees to interact.

We should also note that there are other forms of learning and engagement worth pursuing. Case in point, teams and organizations can devote resources to team learning activities that promote peer learning and consensus-building. Within our own RSE team at Sandia, we hold weekly round-table meetings where team members present on and discuss software engineering best practices\cite{milewicz2020research,willenbring2020,trumbo2021}; we believe this helps cultivate both our team cohesion and expertise in software engineering. Efforts that utilize shared experiences such as these have a low bar to entry and can help build a culture of growth and learning within RSE teams. 

\section{Conclusion}

Across many disciplines, mentorship is considered a vital channel through which workers acquire the mix of skills, knowledge, and sense of professional identity that they need to thrive. In this work, we investigated how mentorship could play an instrumental role in the career growth and retention of RSEs. As research software engineering is an emerging field, however, it is not yet clear what this mentorship should look like or how organizations should support it. To inform the dialogue on this topic, we examined the best available evidence regarding mentorship practices and identified distinct mentorship-related needs that RSEs may have, namely a need for (1) interdisciplinary mentorship networks, (2) long-term mentoring support, and (3) an emphasis on soft skills. Future work in this space could include surveys of the RSE community to identify specific learning needs, experience reports on mentorship within RSE teams, and tailored guidance for national labs and universities looking to implement RSE mentorship programs.

\bibliographystyle{IEEEtran}
\bibliography{rsementorship.bib}

\begin{thebibliography}{10}
\providecommand{\url}[1]{#1}
\csname url@samestyle\endcsname
\providecommand{\newblock}{\relax}
\providecommand{\bibinfo}[2]{#2}
\providecommand{\BIBentrySTDinterwordspacing}{\spaceskip=0pt\relax}
\providecommand{\BIBentryALTinterwordstretchfactor}{4}
\providecommand{\BIBentryALTinterwordspacing}{\spaceskip=\fontdimen2\font plus
\BIBentryALTinterwordstretchfactor\fontdimen3\font minus
  \fontdimen4\font\relax}
\providecommand{\BIBforeignlanguage}[2]{{%
\expandafter\ifx\csname l@#1\endcsname\relax
\typeout{** WARNING: IEEEtran.bst: No hyphenation pattern has been}%
\typeout{** loaded for the language `#1'. Using the pattern for}%
\typeout{** the default language instead.}%
\else
\language=\csname l@#1\endcsname
\fi
#2}}
\providecommand{\BIBdecl}{\relax}
\BIBdecl

\bibitem{baxter2012}
R.~Baxter, N.~C. Hong, D.~Gorissen, J.~Hetherington, and I.~Todorov, ``The
  research software engineer,'' in \emph{Digital Research Conference, Oxford},
  2012, pp. 1--3.

\bibitem{brett2017research}
A.~Brett, M.~Croucher, R.~Haines, S.~Hettrick, J.~Hetherington, M.~Stillwell,
  and C.~Wyatt, ``Research software engineers: state of the nation report
  2017,'' 2017.

\bibitem{cohen2020four}
J.~Cohen, D.~S. Katz, M.~Barker, N.~P.~C. Hong, R.~Haines, and C.~Jay, ``The
  four pillars of research software engineering,'' \emph{IEEE Software}, 2020.

\bibitem{katz2019research}
D.~S. Katz, K.~McHenry, C.~Reinking, and R.~Haines, ``Research software
  development \& management in universities: case studies from manchester's
  rsds group, illinois' ncsa, and notre dame's crc,'' in \emph{2019 IEEE/ACM
  14th International Workshop on Software Engineering for Science
  (SE4Science)}.\hskip 1em plus 0.5em minus 0.4em\relax IEEE, 2019, pp. 17--24.

\bibitem{milewicz2020research}
R.~Milewicz, J.~Willenbring, and D.~Vigil, ``Moving forward together: How a
  software engineering department can impact developer productivity in a
  research organization,'' \emph{Research Software Engineers in HPC
  (RSE-HPC-2020)}, 2020.

\bibitem{billings2020}
J.~Billings, A.~Malviya, and J.~Hendrik, ``Cw20: Jay billings, addi malviya,
  john hendrick interview, rse group, oak ridge national laboratory,''
  \url{https://youtu.be/WBnOLtac4B4}, accessed: 2020-08-07.

\bibitem{deRSEWebsite}
``De-rse,'' \url{https://www.de-rse.org}, accessed: 2021-07-29.

\bibitem{nlRSEWebsite}
``Nl-rse,'' \url{https://nl-rse.org/}, accessed: 2021-07-29.

\bibitem{usSEWebsite}
``Us-rse,'' \url{https://us-rse.org/)}, accessed: 2021-07-29.

\bibitem{nzRSEWebsite}
``Rse-aunz,'' \url{https:// rse-aunz.github.io/}, accessed: 2021-07-29.

\bibitem{ukRSEWebsite}
``Society of research software engineering,'' \url{https://society-rse.org/},
  accessed: 2021-07-29.

\bibitem{se4scienceWorkshops}
``Software engineering for science workshop series,''
  \url{https://se4science.org/workshops/}, accessed: 2021-07-29.

\bibitem{usrseWorkshops}
``Research software engineers in hpc workshop,''
  \url{https://us-rse.org/rse-hpc-2020/}, accessed: 2021-07-29.

\bibitem{rseconWorkshops}
``The fourth conference of research software engineers,''
  \url{https://rse.ac.uk/conf2019/}, accessed: 2021-07-29.

\bibitem{hobson2009mentoring}
A.~J. Hobson, P.~Ashby, A.~Malderez, and P.~D. Tomlinson, ``Mentoring beginning
  teachers: What we know and what we don't,'' \emph{Teaching and teacher
  education}, vol.~25, no.~1, pp. 207--216, 2009.

\bibitem{sambunjak2010systematic}
D.~Sambunjak, S.~E. Straus, and A.~Marusic, ``A systematic review of
  qualitative research on the meaning and characteristics of mentoring in
  academic medicine,'' \emph{Journal of general internal medicine}, vol.~25,
  no.~1, pp. 72--78, 2010.

\bibitem{andrews1999mentorship}
M.~Andrews and M.~Wallis, ``Mentorship in nursing: a literature review,''
  \emph{Journal of advanced nursing}, vol.~29, no.~1, pp. 201--207, 1999.

\bibitem{trainer2017mentoring}
E.~H. Trainer, A.~Kalyanasundaram, and J.~D. Herbsleb, ``E-mentoring for
  software engineering: A socio-technical perspective,'' in \emph{2017 IEEE/ACM
  39th International Conference on Software Engineering: Software Engineering
  Education and Training Track (ICSE-SEET)}.\hskip 1em plus 0.5em minus
  0.4em\relax IEEE, 2017, pp. 107--116.

\bibitem{canfora2012going}
G.~Canfora, M.~Di~Penta, R.~Oliveto, and S.~Panichella, ``Who is going to
  mentor newcomers in open source projects?'' in \emph{Proceedings of the ACM
  SIGSOFT 20th International Symposium on the Foundations of Software
  Engineering}, 2012, pp. 1--11.

\bibitem{fagerholm2014role}
F.~Fagerholm, A.~S. Guinea, J.~M{\"u}nch, and J.~Borenstein, ``The role of
  mentoring and project characteristics for onboarding in open source software
  projects,'' in \emph{Proceedings of the 8th ACM/IEEE international symposium
  on empirical software engineering and measurement}, 2014, pp. 1--10.

\bibitem{lienard2018intellectual}
J.~F. Li{\'e}nard, T.~Achakulvisut, D.~E. Acuna, and S.~V. David,
  ``Intellectual synthesis in mentorship determines success in academic
  careers,'' \emph{Nature communications}, vol.~9, no.~1, pp. 1--13, 2018.

\bibitem{malmgren2010role}
R.~D. Malmgren, J.~M. Ottino, and L.~A.~N. Amaral, ``The role of mentorship in
  prot{\'e}g{\'e} performance,'' \emph{Nature}, vol. 465, no. 7298, pp.
  622--626, 2010.

\bibitem{kram1983phases}
K.~E. Kram, ``Phases of the mentor relationship,'' \emph{Academy of Management
  journal}, vol.~26, no.~4, pp. 608--625, 1983.

\bibitem{wright1997perceptions}
N.~Wright and M.~Bottery, ``Perceptions of professionalism by the mentors of
  student teachers,'' \emph{Journal of Education for Teaching}, vol.~23, no.~3,
  pp. 235--252, 1997.

\bibitem{jazayeri2004}
M.~Jazayeri, ``The education of a software engineer,'' in \emph{Proceedings.
  19th International Conference on Automated Software Engineering, 2004.}\hskip
  1em plus 0.5em minus 0.4em\relax IEEE, 2004, pp. xviii--xxvii.

\bibitem{de2003}
S.~C. De~Janasz, S.~E. Sullivan, and V.~Whiting, ``Mentor networks and career
  success: Lessons for turbulent times,'' \emph{Academy of Management
  Perspectives}, vol.~17, no.~4, pp. 78--91, 2003.

\bibitem{henry2019mentorship}
N.~Henry-Noel, M.~Bishop, C.~K. Gwede, E.~Petkova, and E.~Szumacher,
  ``Mentorship in medicine and other health professions,'' \emph{Journal of
  Cancer Education}, vol.~34, no.~4, pp. 629--637, 2019.

\bibitem{guise2017team}
J.-M. Guise, S.~Geller, J.~G. Regensteiner, N.~Raymond, and J.~Nagel, ``Team
  mentoring for interdisciplinary team science: lessons from k12 scholars and
  directors,'' \emph{Academic medicine: journal of the Association of American
  Medical Colleges}, vol.~92, no.~2, p. 214, 2017.

\bibitem{freel2017multidisciplinary}
S.~A. Freel, P.~C. Smith, E.~N. Burns, J.~B. Downer, A.~J. Brown, and M.~W.
  Dewhirst, ``Multidisciplinary mentoring programs to enhance junior faculty
  research grant success,'' \emph{Academic medicine: journal of the Association
  of American Medical Colleges}, vol.~92, no.~10, p. 1410, 2017.

\bibitem{kensington2018surviving}
B.~Kensington-Miller, ``Surviving the first year: new academics flourishing in
  a multidisciplinary community of practice with peer mentoring,''
  \emph{Professional development in education}, vol.~44, no.~5, pp. 678--689,
  2018.

\bibitem{stein2010}
M.~Stein and L.~Christiansen, \emph{Successful onboarding}.\hskip 1em plus
  0.5em minus 0.4em\relax McGraw-Hill Professional Publishing, 2010.

\bibitem{dice2016}
\BIBentryALTinterwordspacing
D.~Insights. (2016) How long do tech pros stay in their jobs? [Online].
  Available:
  \url{https://insights.dice.com/2016/07/08/how-long-do-tech-pros-stay-in-their-jobs/}
\BIBentrySTDinterwordspacing

\bibitem{matturro2015soft}
G.~Matturro, F.~Raschetti, and C.~Font{\'a}n, ``Soft skills in software
  development teams: A survey of the points of view of team leaders and team
  members,'' in \emph{2015 IEEE/ACM 8th International Workshop on Cooperative
  and Human Aspects of Software Engineering}.\hskip 1em plus 0.5em minus
  0.4em\relax IEEE, 2015, pp. 101--104.

\bibitem{capretz2018call}
L.~F. Capretz and F.~Ahmed, ``A call to promote soft skills in software
  engineering,'' \emph{Psychology and Cognitive Sciences}, 2018.

\bibitem{matturro2019systematic}
G.~Matturro, F.~Raschetti, and C.~Font{\'a}n, ``A systematic mapping study on
  soft skills in software engineering,'' \emph{Journal of Universal Computer
  Science}, vol.~25, no.~1, pp. 16--41, 2019.

\bibitem{van2018survey}
B.~Van~Werkhoven, T.~Bakker, O.~Philippe, and S.~Hettrick, ``Survey on research
  software engineering in the netherlands,'' in \emph{2018 IEEE 14th
  International Conference on e-Science (e-Science)}.\hskip 1em plus 0.5em
  minus 0.4em\relax IEEE, 2018, pp. 38--39.

\bibitem{willenbring2020}
J.~Willenbring and R.~Milewicz, ``Moving forward together: How a software
  engineering department can impact developer productivity in a research
  organization,'' \emph{2020 Collegeville Workshop on Scientific Software},
  2020.

\bibitem{trumbo2021}
D.~Trumbo and R.~Milewicz, ``Towards a culture of continuous improvement within
  rse teams,'' in \emph{US-RSE Virtual Workshop 2021}.\hskip 1em plus 0.5em
  minus 0.4em\relax US-RSE, 2021.

\end{thebibliography}

\end{document}